\documentclass[journal=aamick,manuscript=letter,layout=twocolumn]{achemso}

\usepackage[version=3]{mhchem} 
\usepackage{graphicx}
\usepackage{dcolumn}
\usepackage{bm}
\usepackage[mathlines]{lineno}
\usepackage{amssymb}
\usepackage{amsmath}
\usepackage{soul,color}
\usepackage[colorlinks,linkcolor=blue,anchorcolor=blue,citecolor=blue,urlcolor=blue]{hyperref}
\usepackage{dcolumn}

\usepackage{threeparttable}  
\usepackage{cmll}
\usepackage{multirow}
\usepackage{booktabs}
\usepackage{color}

\title{Insight into Two-Dimensional Borophene: Five-Center Bond and Phonon-Mediated Superconductivity}

\author{Zhibin Gao}
\email{zhibin.gao@nus.edu.sg}
\affiliation{Department of Physics, National University of Singapore,
             Singapore 117551, Republic of Singapore}
\author{Mengyang Li}
\affiliation{Institute for Chemical Physics $\with$ Department of Chemistry, 
             Graduate School of Science, Xi'an Jiaotong University,
             Xi'an 710049, China}
\author{Jian-Sheng Wang}
\affiliation{Department of Physics, National University of Singapore,
             Singapore 117551, Republic of Singapore}

\date{\today}

\begin{document}


\begin{abstract}
%
%
We report a previously unknown monolayer borophene allotrope and we call
it super-B with a flat structure based on the $\it{ab~initio}$ calculations.
It has good thermal, dynamical, and mechanical stability compared with
many other typical borophenes. We find that super-B has a fascinating
chemical bond environment consisting of standard \textit{sp},
\textit{sp$^2$} hybridizations and delocalized five-center three-electron
$\pi$ bond, called $\pi$(5c-3e). This particular electronic structure
plays a pivotal role in stabilizing the super-B chemically.
By extra doping, super-B can be transformed into a Dirac material from
pristine metal. Like graphene, it can also sustain tensile strain smaller
than 24\%, indicating superior flexibility. Moreover, due to
the small atomic mass and large density of states at the Fermi level, super-B
has the highest critical temperature \textit{T$_c$} of 25.3~K in single-element
superconductors at ambient condition. We attribute this high \textit{T$_c$} of
super-B to the giant anharmonicity of two linear acoustic phonon branches
and an unusually low optic phonon mode. These predictions provide
new insight into the chemical nature of low dimensional boron nanostructures
and highlight the potential applications of designing flexible
devices and high \textit{T$_c$} superconductor.
%
\end{abstract}

\section*{Introduction}
Boron atom has five electrons with an electron
configuration 1s$^2$2s$^2$2p$^1$,
which means that the number
of valence electrons is less than the available orbitals. Due to
the small atomic radius, removal of valence electrons of boron
requires a large amount of energy. Therefore, boron likely forms
covalent compounds, rather than B$^{3+}$ ions. However, owing to
the smaller electronegativity than hydrogen, boron atoms have a
little positive charge in most covalent compounds,  except for
special B-B bonds.




Matter always wants to be in the most stable form. For most of non-metal
atoms, stability is achieved by following the octet rule. Theoretically,
boron can accommodate five more electrons according to the octet rule,
%
but boron commonly emerges
three bonds, like BH$_3$, with a total of six electrons in the outermost
shell excepting for coordination compounds with four ligands like
[BF$_4$]$^-$ and B[OH]$^-_4$.
Hence, conventional 2c-2e bond
does not hold in the ``electron-deficiency"
boron compounds\cite{li2017planar,sergeeva2014understanding,wang2016photoelectron}.
Afterwards, many researchers find that localized three-centered bonds,
multi-centered bonds, and delocalized $\sigma$ and $\pi$ bonds play very
crucial roles in stabilizing the boron
compounds\cite{eberhardt1954valence,lipscomb1977boranes}. This
particular electronic configuration also results in many
anomalous properties in boron
compounds\cite{nagamatsu2001superconductivity,feng2017dirac,li2017planar}.

In 2012, for the first time, polymorphism of 2D boron, called
borophene, was proposed\cite{penev2012polymorphism} and proper substrates
were further explored theoretically\cite{liu2013probing,zhang2015two}, which
play pivotal roles in finally synthesized borophene
experimentally\cite{mannix2015synthesis,feng2016experimental}.

In this study, we predict a previously unknown 2D borophene allotrope,
called super-B due to the large vacancy, based on the \textit{ab initio}
calculations. Super-B is a hexagonal borophene with three atoms in
each side, different from the hexagonal graphene.
Besides, it has good thermal, dynamical, and mechanical stability. We find that
the structure of super-B is formed by the standard \textit{sp}, \textit{sp$^2$}
and delocalized $\pi$(5c-3e) bonds in terms of the natural bond
orbital (NBO) analysis. These particular type of chemical bonds have never
been reported in borophene before. Therefore, super-B sets
a good example to break the stereotype that triangular lattice with hexagonal
vacancies is the only principle and concept of borophene.
Furthermore, by doping method,
super-B can be transformed into a Dirac material from pristine metal.
It can also sustain 24\% of biaxial strain before fracture, indicating
the same superior flexibility as graphene. Due to the small atomic mass
and large density of states at Fermi level, super-B has a phonon-mediated
superconducting \textit{T$_c$} of 25.3~K, which is derived from the
giant anharmonicity of two linear acoustic phonon branches and an
unusually low optic \textit{O$_z$} phonon mode.

\section*{Computational details}
We studied the electronic structure, the equilibrium geometry, and
structural stability of super-B using \textit{ab initio} DFT as
implemented in VASP\cite{kresse1996efficient,kresse1996efficiency,kresse1994ab}.
We used the exchange-correlation functional of
Perdew-Burke-Ernzerhof (PBE)\cite{perdew1996generalized} and
HSE06\cite{heyd2003hybrid,krukau2006influence} with default mixing
parameter value $\alpha$ = 0.25. The vacuum distance over adjacent
boron layer is set to be 20 \AA. The plane-wave cutoff is set to 500~eV.
For structure optimization, both lattice constants and atomic positions
are relaxed with the criterion for total energy of 1.0 $\times$ 10$^{-8}$ eV
and Hellmann-Feynman forces of 10$^{-4}$ eV/ \AA. The Brilliouin zone is
sampled by 11 $\times$ 11. For the electron-phonon coupling (EPC), we used
Quantum Espresso\cite{giannozzi2009quantum} with 80~Ry energy cutoff.
The self-consistent electron density in super-B was calculated by a
64 $\times$ 64 k-point grid. The dynamical matrices were calculated on a
10 $\times$ 10 grid
after convergence test. The NBO analysis
was carried out on B3LYP/6-311G(d,p) with Gaussian 16\cite{frisch2016gaussian}
after optimization of B3LYP/6-31G(d,p).

\section*{Results}
\subsection*{Structure}

\begin{figure*}[t!]
\includegraphics[width=1.7\columnwidth]{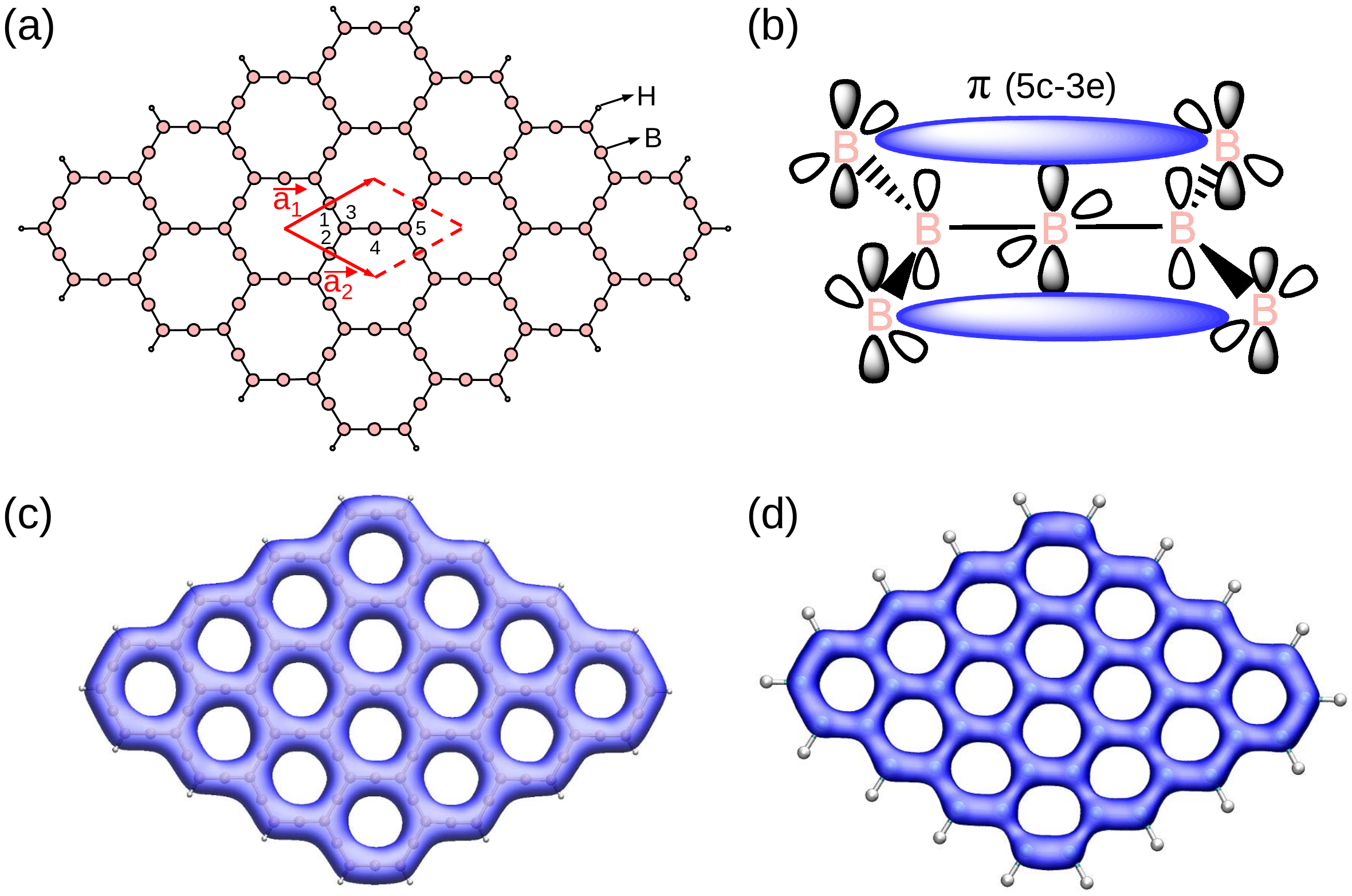}
\caption{(a) Crystal structure of super-borophene (super-B). The
marginal boron atoms are saturated by hydrogens and the primitive
cell is denoted by a red box. Each boron atom is labelled by a cardinal
number. (b) The hybridized orbital analysis in the primitive cell.
Dotted and solid wedge shapes represent the boron atoms perpendicular
to the plane of paper outside and inside respectively, which means
seven boron atoms stand in same plane. One \textit{p} orbital of
each boron atom, totally seven, is perpendicular to plane of super-B
(parallel to plane of paper), which is shown by two shaded dumbbells
with one electron for \textit{sp} and two white dumbbells without electron
for \textit{sp$^2$} hybridization. All delocalized orbitals
perpendicular to the plane of super-B construct the large $\pi$(5c-3e) bond.
The other \textit{p} orbitals of \textit{sp} boron atoms is parallel
to the plane of super-B shown by two white dumbbells. Localized
orbital locator (LOL) maps for (c) super-B and (d) graphene, showing
strong $\pi$ bonds (isovalue = 0.4 a.u.).}\label{fig1}
\end{figure*}

\begin{figure*}[t!]
\includegraphics[width=1.7\columnwidth]{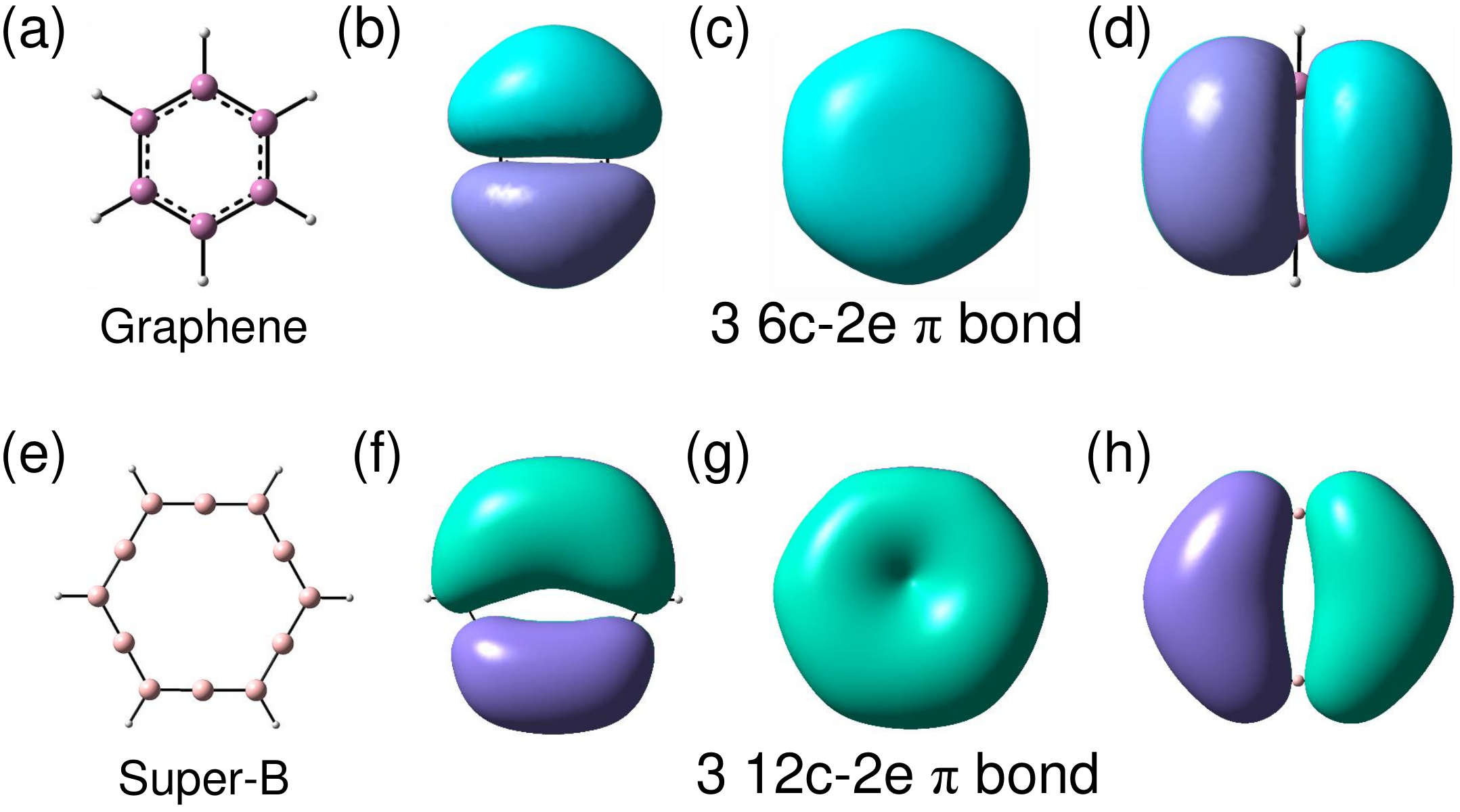}
\caption{(a)(e) Optimized structures of benzene (C$_6$H$_6$) and
hydrogenated super-B (B$_{12}$H$_6$). Molecular orbitals of
(b)(c)(d) 6c-2e $\pi$ bonds of benzene and (f)(g)(h) 12c-2e $\pi$
bonds of hydrogenated super-B based on the AdNDP
method\cite{zubarev2008developing} (white ball stands for hydrogen).}
\label{fig2}
\end{figure*}

The optimized super-B is shown in Figure~\ref{fig1}a.
The lattice constants are
$\left|\vec{a_1}\right|$ = $\left|\vec{a_2}\right|$ = 5.45 {\AA} with
\textit{P}6/\textit{mmm} (space group no. 191). There are two independent
atoms based on the group Wyckoff positions, whose fractional coordinates
are ($\frac{1} {3}$, $\frac{2} {3}$) and ($\frac{1} {2}$, $\frac{1} {2}$).
In order to perform the chemical bond analysis quantitatively,
we use hydrogen atoms to saturate boron atoms at the border and explore
the primitive cell located in the center region marked by a red
box with negligible fringe effect. Each boron atom is labelled
by a subscript number. We find that B$_3$ is coordinated with
three B (B$_1$, B$_2$ and B$_4$) via \textit{sp$^2$} hybridization,
and B$_4$ is linked to the B$_3$ and B$_5$ with \textit{sp}
hybridization. The detailed NBO analysis is shown in Table ~\ref{Tab1}.
It is clear that each boron atom has a \textit{p} orbital which is
perpendicular to the plane of super-B in Figure~\ref{fig1}b, and
these \textit{p} orbitals of boron, unhybridized and perpendicular
to the plane of super-B, constitute an electron-deficiency
$\pi$(5c-3e) bond in the primitive cell, which is
similar to the $\pi$(2c-2e) orbital in the primitive cell of graphene.
Therefore, on the basis of periodicity rule, a large delocalized
$\pi$(5nc-3ne) bond with electron-deficiency characteristic emerges
on the surface of super-B.

In order to further verify the interesting $\pi$ bond,
localized orbital locator (LOL) maps in Figure~\ref{fig1}c and
Figure~\ref{fig1}d of super-B and graphene with isovalue 0.4 a.u.
have been conducted by Multiwfn~\cite{schmider2000chemical,lu2012multiwfn}.
All the $\pi$ bonds are well depicted in LOL picture.  It is evident
that the $\pi$ electrons from \textit{p} orbitals of \textit{sp}
boron delocalized on the whole plane of super-B. All these
delocalized large $\pi$ bonds, analogous with graphene,
significantly contribute to the stability of
super-B.
Interestingly, the chemical bonds of super-B with $\pi$(5c-3e)
bond have not been reported before in 2D materials and are quite
different from the conventional bond in borophene allotrope,
like three-centered bond in triangular lattice with/without
vacancies\cite{tang2007novel,penev2012polymorphism}.

In addition, the similarity of delocalized large $\pi$
bonds between graphene and super-B is also reflected by the electronic
configurations shown in Figure~\ref{fig2}. The relationship between
benzene (C$_6$H$_6$) and graphene is the same as hydrogenated
super-B (B$_{12}$H$_6$) and super-B. The electronic structure of
benzene is well-known as three completely delocalized $\pi$ orbitals,
6c-2e $\pi$ bonds\cite{popov2012chemical}, shown in
Figure~\ref{fig2}b-2d based on the adaptive nature
density partitioning (AdNDP)\cite{zubarev2008developing}.
B$_{12}$H$_6$ is composed of three completely delocalized 12c-2e
$\pi$ bonds with a little electron-deficiency characteristic. Here,
the completely delocalized $\pi$ bond in super-B is verified again.
Besides, we also find that there is no Jahn-Teller distortion in
super-B by optimizing the atomic positions in a large supercell.

\begin{table*}
\scriptsize
\centering
\renewcommand\arraystretch{1.5}
\caption{The natural bond orbital (NBO) analysis of atomic notation
B$_3$ and B$_4$ in Figure~\ref{fig1}a including the number of
occupancy with spin-up, spin-down electrons of boron and total
of them, as well as contributions of \textit{s}, \textit{p} and
\textit{d} atomic orbitals.}\label{Tab1}
\setlength{\tabcolsep}{4.5mm}{
\begin{tabular}{cccccccccc}
\hline \hline
\multirow{2.30}{*}{Center atoms} & \multirow{2.30}{*}{Bonding atoms} & \multicolumn{3}{c}{Occupancy} & \multicolumn{4}{c}{Hybridization of center boron$^a$}  \\
\cmidrule(r){3-5} \cmidrule(r){6-9}
& &  Spin-up   &  Spin-down   &   total
&  \textit{$sp^\lambda$}   &  \textit{s(\%)}      &  \textit{p(\%)}   &   \textit{d(\%)}  \\
\midrule
B$_3$  & B$_1$  & 0.95  & 0.95  & 1.90   & \textit{$sp^{1.99}$}  & 33.37  & 66.53  & 0.10    \\
B$_3$  & B$_2$  & 0.95  & 0.95  & 1.90   & \textit{$sp^{1.99}$}  & 33.37  & 66.53  & 0.10    \\
B$_3$  & B$_4$  & 0.95  & 0.95  & 1.90   & \textit{$sp^{2.02}$}  & 33.06  & 66.84  & 0.10    \\
B$_4$  & B$_3$  & 0.95  & 0.95  & 1.90   & \textit{$sp^{1.00}$}  & 49.99  & 50.00  & 0.01    \\
B$_4$  & B$_5$  & 0.95  & 0.95  & 1.90   & \textit{$sp^{1.00}$}  & 49.99  & 50.00  & 0.01    \\
\hline \hline
\end{tabular}}
 \begin{tablenotes}
        \footnotesize
        \item[a]$^a$The contribution of \textit{s}, \textit{p} and \textit{d} orbitals
                of center boron is the average of spin-up, spin-down electrons of
                boron, and \textit{$sp^\lambda$} of center boron is the almost same
                for both spin-up and spin-down.
 \end{tablenotes}
\end{table*}

\begin{table*}
 \begin{threeparttable}
		\caption{The calculated lattice constants, planer or buckling, coordination
                 number, symmetry and cohesive energy for 2D borophene allotropes
                 $\chi_3$, $\beta_{12}$, $\delta_4$, $\delta_6$ and our proposed
                 super-B.}\label{Tab2}
		\renewcommand\arraystretch{1.5}
		\begin{tabular*}{1.0\textwidth}{p{2.7cm}p{1.6cm}*{2}{p{0.10\textwidth}}p{1.8cm}p{2.5cm}p{2.5cm}}		
			\hline \hline
	        Structures  & $a_1$ (\AA)  & $a_2$ (\AA) & Planer & ~\textit{Z} & Symmetry & \textit{E$_c$} (eV/atom)   \\
			\hline
			$\chi_3$-B$^a$      & 4.45   & 4.45   & ~Yes    & 4,5       & \textit{C}mmm	     & ~~~5.723		 \\
			$\beta_{12}$-B$^a$  & 2.92   & 5.07   & ~Yes    & 4,5,6     & \textit{P}mmm      & ~~~5.712    	 \\
            $\delta_4$-B$^a$    & 2.93   & 3.28   & ~Yes    & 4         & \textit{P}mmm      & ~~~5.384      \\
			$\delta_6$-B$^a$  	& 3.22   & 3.29   & ~No     & 6         & \textit{P}mmn      & ~~~5.662       \\
			Super-B$^b$  	    & 5.45   & 5.45   & ~Yes    & 2,3       & \textit{P}6/mmm    & ~~~5.550       \\
			\hline \hline
		\end{tabular*}
 \begin{tablenotes}
        \footnotesize
        \item[]$^a$Ref~\cite{wu2012two,feng2016experimental}. $^{b}$Present work. All results are based on
        the PBE functionals.
      \end{tablenotes}
    \end{threeparttable}
\end{table*}

\begin{figure*}[t!]
\includegraphics[width=1.9\columnwidth]{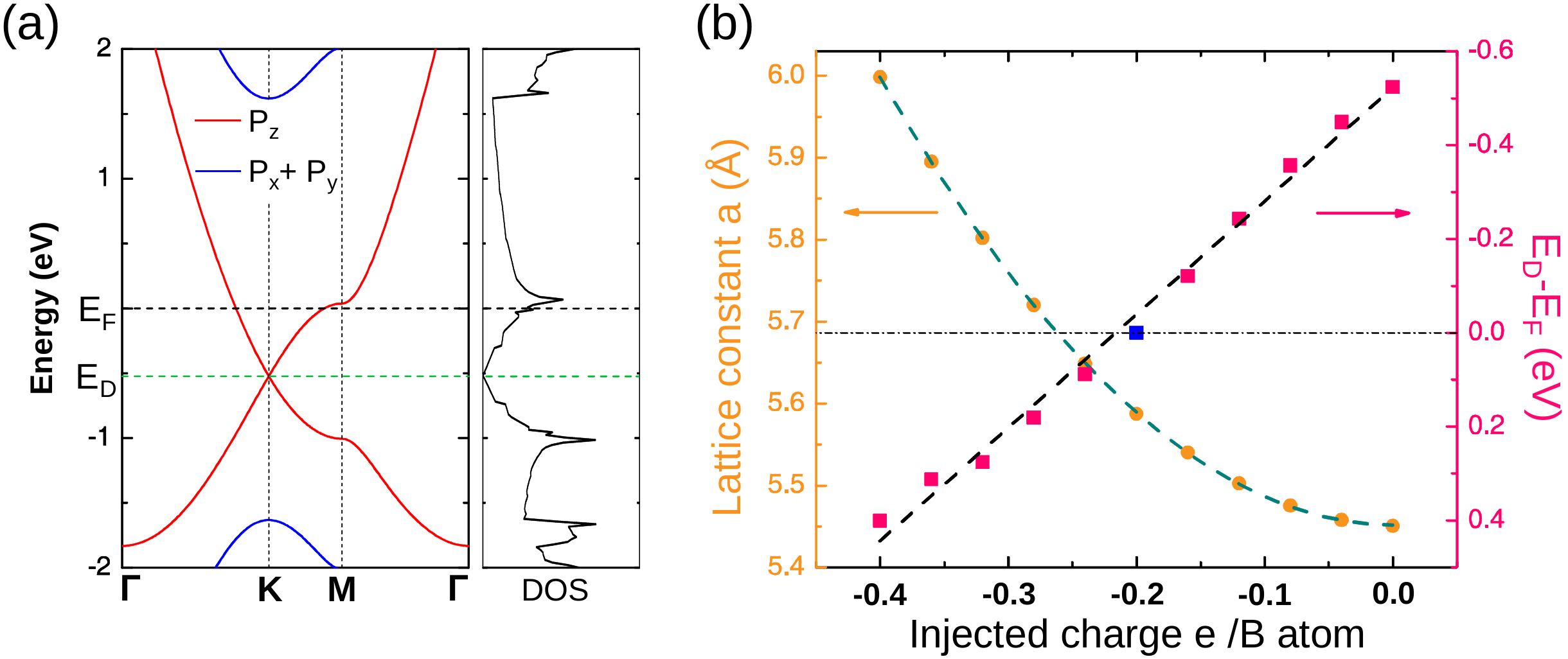}
\caption{(a) Orbitally resolved band structure and total
density of states for pristine super-borophene. (b) The
lattice constant and the position of Dirac cone (\textit{E$_D$})
with respect to the Fermi level (\textit{E$_F$}) as a function
of injected charge. The green dashed line is a parabolic
fitting of lattice constant with 0.999 R-Squared and the black
dashed line is a linear fitting of the relative gap
between \textit{E$_F$} and \textit{E$_D$} with 0.992 R-Squared.
Blue square denotes the situation of \textit{E$_F$} = \textit{E$_D$}.
The dashed horizontal lines are guided for your eyes.}\label{fig3}
\end{figure*}

The cohesive energy \textit{E$_c$} is the amount of energy to break
a material into isolated atoms. Calculated \textit{E$_c$} of super-B
is 5.55~eV/atom, only 0.1~eV/atom higher than
$\delta_6$-B\cite{feng2016experimental}, but more stable than the
$\delta_4$-B\cite{wu2012two} shown in Table ~\ref{Tab2}. For any
mechanically stable 2D materials, a necessary, but not a sufficient
condition must be satisfied:
C$_{11}$C$_{22}$-C$_{12}^{2}$$>$0 and C$_{66}$$>$0\cite{gao2017novel}.
The calculated elements of super-B are C$_{11}$=C$_{22}$=146.0~GPa,
C$_{12}$=138.1~GPa, and C$_{66}$=3.9~GPa, assuming an effective
thickness of 0.384~nm (two van der Waals radius), which verifies
the mechanical stability of super-B. The phonon dispersion is shown in the
Supporting Information. Obviously, there are two linear LA and TA acoustic
phonon branches, and a parabolic ZA around $\Gamma$-point. All frequencies
are free from imaginary, confirming the lattice dynamical stability. Besides,
we also implemented  \textit{ab initio} molecular dynamics simulations
using canonical ensemble and Nos\'{e}-Hoover thermostat at 600~K for
10~ps. The movie, shown in the Supporting Information, indicates that
the robustness of our predicted super-B.
Besides, we find that super-B is non-magnetic, even
having an odd number of electrons based on the first-principle calculation.
Therefore, our calculation
confirms that super-B is chemically, thermodynamically, mechanically, and
dynamically stable.

\begin{figure*}[t!]
\includegraphics[width=1.8\columnwidth]{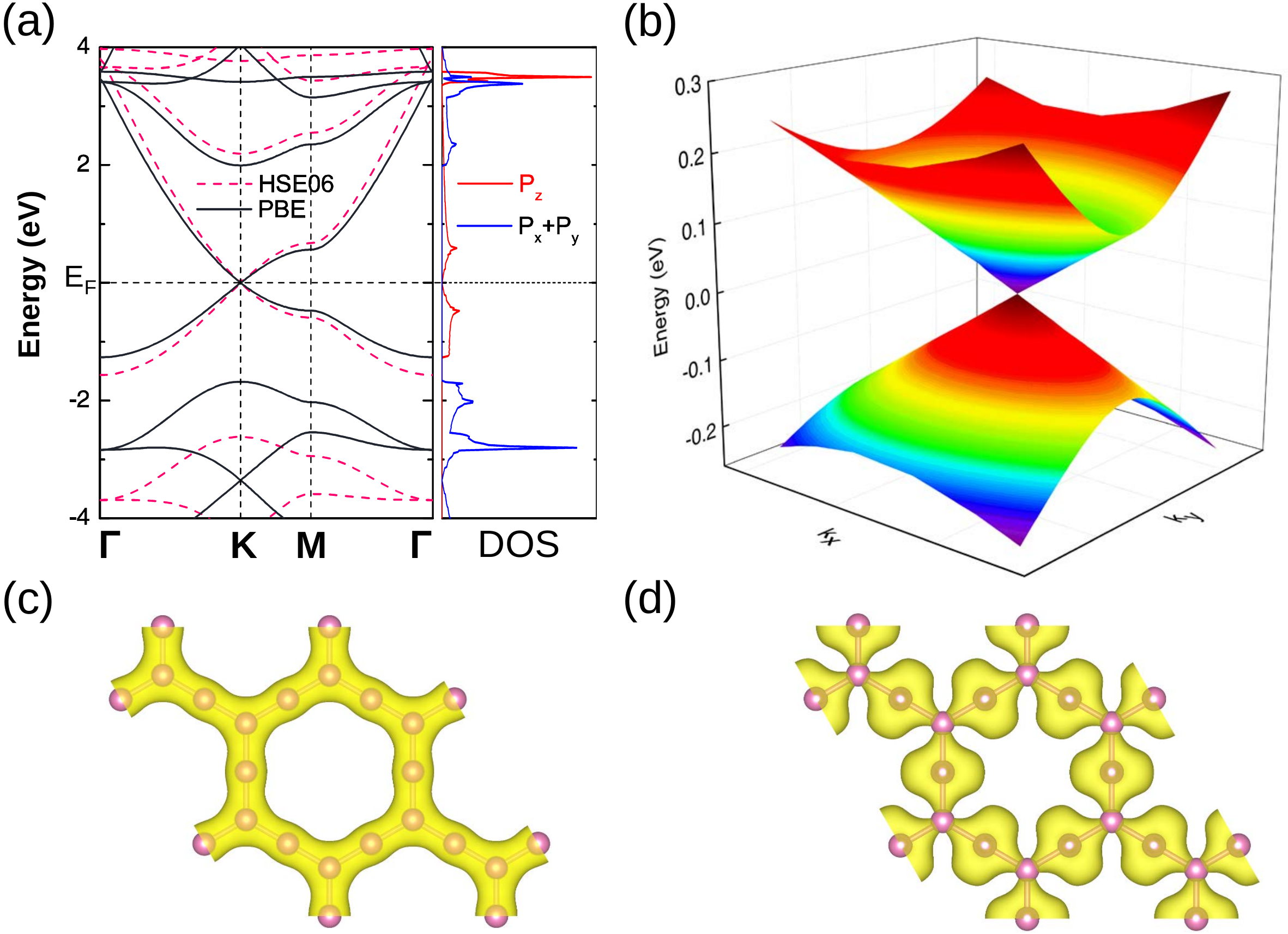}
\caption{(a) Electronic band structures from PBE (black solid line)
and HSE06 (magenta dashed line) methods and projected density of
states (PBE level) of super-B with injected electrons
$\left \langle \Delta Q \right \rangle$ = $- 0.2$~e/B, corresponding
to a doping level of 7.7 $\times$ 10$^{13}$ cm$^{-2}$ that has been
realized by electrical gating
and ionic liquid injection.
(b) 3D Dirac cone formed by the valence and conduction bands in the
vicinity of Dirac point. In the first Brillouin zone, the high symmetry
$k$ points are: $\Gamma$ (0 0 0), K (-1/3 2/3 0), and M (0 0.5 0).
(c,d) The isosurfaces of partial charge densities for the (c) VBM
and (d) CBM of super-B with
$\left \langle \Delta Q \right \rangle$ = $-0.2$e/B. The isosurface
level value is 0.03 e {\AA}$^{-3}$.}\label{fig4}
\end{figure*}

Intriguingly, we find a similar pattern, called $\alpha$-graphyne, has
been explored in the carbon counterpart but is not dynamically
stable\cite{longuinhos2014}. However, inserting two threefold-coordinated
carbon atoms into a 2D hexagonal lattice makes $\beta$-graphyne
stable\cite{malko2012competition}. We did a similar analysis and found
that the phonon dispersion of $\beta$-borophyne has large imaginary
frequency, reflecting the significant distinction of electron orbitals
between carbon and boron. Considering $\beta$-graphyne substructures
and derivatives that have been experimentally synthesized in large-area
nano-films\cite{li2010architecture,inagaki2014graphene}, we are
confident that the same experimental technique can be extended
to the super-B and boron nanostructures on the horizon.

\subsection*{Electronic structure and doping effect}
The projected electronic band structure and density of states (DOS)
of super-B are shown in Figure~\ref{fig3}a. Like graphene,
\textit{p$_z$} orbital entirely dominates the electronic behavior
around the Fermi level. We find that the intrinsic super-B is metallic,
which is analogous with $\alpha$-B\cite{tang2007novel},
$\beta_{12}$-B\cite{feng2016experimental}, and other monolayer boron
sheets\cite{penev2012polymorphism,li2017planar}. Below the \textit{E$_F$}
around 0.52~eV, two \textit{p$_z$} bands cross each other, forming
a standard Dirac cone (\textit{E$_D$}), which is verified by the
total DOS in the right panel.

Recently, honeycomb borophene by extra doping of 1~e/B atom, identical
with graphene's configuration, has been realized
experimentally\cite{li2018experimental} and
theoretically\cite{liu2019effect}.
Here, we would like to explore how
large doping could make super-B a semimetal, like graphene. Since
\textit{E$_F$} is larger than \textit{E$_D$} in the original state,
one should provide a degree of hole doping to the super-B, shifting
up the bands. The results are shown in Figure~\ref{fig3}b. In the
calculation, when adding or removing an amount of electrons (holes),
the system will be compensated by the same amount uniform background
charge of holes (electrons), retaining a continuous neutral condition.
This standard method has been verified by many previous
works\cite{liu2019effect}.

In each doping concentration, we optimized both lattice constants and
atomic positions. On the one hand, the lattice constant of super-B,
interestingly, can be well fitted by a quadratic polynomial with
0.999~R-Squared as a function of hole doping. Due to the special
\textit{sp}, \textit{sp$^2$} and $\pi$(5c-3e) bond, the bond
length increases from 1.573~\AA~to 1.613~\AA~when injecting $-0.2$/B
and further reaches to 1.731~\AA~at $-0.4$/B. This phenomenon is
different from the previous orthogonal $\epsilon$-B
allotrope\cite{liu2019effect} in which doping has an anisotropic
effect on the lattice constants. Due to the \textit{P}6/mmm
symmetry of super-B, the net charge has the same effect on both a$_1$
and a$_2$ lengths. In this sense, doping is an effective way to
change the lattice constant of super-B and other 2D
borophenes\cite{liu2019effect,si2012electronic}.

On the other hand, doping will change the electronic band structure,
especially the position of \textit{E$_F$}. The relative position of
\textit{E$_D$} with respect to \textit{E$_F$}, in Figure~\ref{fig3}b,
decreases when increasing the injected holes. We find that the
\textit{E$_D$}-\textit{E$_F$} can be well fitted by a linear function
with 0.992~R-Squared, indicating a rigid band shift in super-B.
$-0.4$/B doping corresponds to 1.44 $\times$ 10$^{14}$ cm$^{-2}$, which
is an accessible and reachable value in the current experimental technique,
such as electrical gating and ionic liquid injection. Even though the
rigid band shift is common in 2D transition metal
dichalcogenides\cite{gao2019degenerately}, we have not found a similar
behavior in 2D boron allotrope heretofore. The reason behind it is that
many 2D borophenes mainly possess three-centered bond
characteristic\cite{tang2007novel,penev2012polymorphism}, rather
than \textit{sp}, \textit{sp$^2$}, and delocalized $\pi$(5c-3e) bond
in super-B.

\begin{figure*}[t!]
\includegraphics[width=2.0\columnwidth]{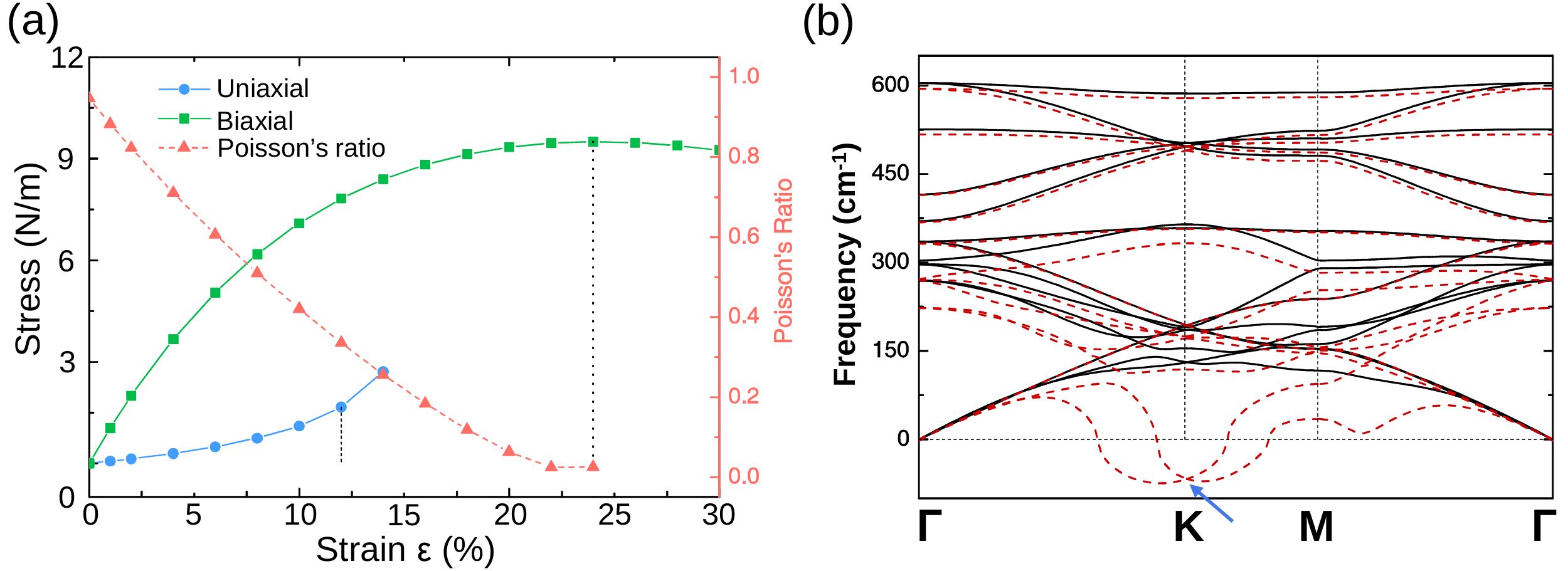}
\caption{(a) Strain-stress curves of super-borophene while
loaded along biaxial (green) and uniaxial (blue) directions,
and Poisson's ratio (pink) under biaxial strain. (b) phonon dispersion
under 24\% (solid black line) and 25\% (red dashed line) biaxial
strain. The critical points of both types of strain are marked by
vertical dashed lines in (a). The blue arrows illustrates
the ``Kohn anomaly" related to lattice instability.}\label{fig5}
\end{figure*}

In order to further confirm the Dirac cone of super-B at
$-0.2$/B (7.7 $\times$ 10$^{13}$ cm$^{-2}$) doping concentration, we plot
the electronic band and corresponding DOS in Figure~\ref{fig4}a using
PBE (black solid line) and standard HSE06 methods (pink dashed line). It
shows that \textit{p$_x$}+\textit{p$_y$} shift up or down using HSE06.
However, the bands from \textit{p$_z$} orbital do not change its
position, only having slightly different slopes of two linear bands when
hybrid functional is applied. 3D band in the reciprocal space, in
Figure~\ref{fig4}b, further proves the band crossing at the K-point.
Besides, we compare the band structures of $\alpha$-graphyne and hole-doped
super-B in the Supporting Information. They look similar but they have
different Fermi velocities $\upsilon_F$. Even though $\upsilon_F$ of super-B
is smaller than $\alpha$-graphyne, we show the possibility to transform from
metal to semimetal in 2D boron system by extra doping. Furthermore, we
calculate the partial charge densities of valence
band maximum (VBM) and conduction band minimum (CBM) of super-B in
Figure~\ref{fig4}c and 4d. VBM stems from the delocalized orbitals, resulting in
large $\pi$(5c-3e) bond that has mentioned in Figure~\ref{fig1}c,
whereas CBM shows localized in-plane orbitals constituting the standard
\textit{sp} and \textit{sp$^2$} bonds between boron atoms. This picture is
well consistent with the chemical NBO analysis shown in Figure~\ref{fig1}.

\subsection*{Mechanical property}
In practical applications, a large ideal strength of material is highly
desired in flexible electronic
devices\cite{zhang2017elasticity,liu1989prediction,si2013first}. According
to the definition, biaxial tensile strain can be expressed as
$\varepsilon = a/a_0 - 1$, in which $a$ and $a_0$ are the stretched and
pristine lattice constant of materials. As biaxial strain $\varepsilon$
increases, in Figure~\ref{fig5}a, the stress $\sigma$ in green line
first linearly increases, then gradually saturates with a maximum value
called ideal strength\cite{gao2017novel}. The ideal strength of super-B
is 9.50~N/m,
which is the same order of many other
borophenes\cite{zhang2017elasticity,wang2016strain}. In each strain, we
calculate the phonon dispersion to verify its stability. Smaller than
$\varepsilon \leq$ 24\%, all phonon frequencies are positive, shown in
Figure~\ref{fig5}b (black line). When $\varepsilon$=25\%, the system
becomes unstable, which is a clear hint of the ``Kohn anomaly''
when reaching the ideal strength in super-B. $\varepsilon$ = 25\% strain is
also called breaking point which means a material physically breaks at its
breaking point.
This phenomenon has been studied in graphene
extensively\cite{yan2008observation,lazzeri2006nonadiabatic,si2012electronic}.

For the uniaxial strain, the isotropically stress response of super-B is
shown in Figure~\ref{fig5}a in blue line. The ideal strength is 1.66~N/m
at $\varepsilon$=12\%, relatively smaller than the biaxial strain (phonon
dispersion is shown in the Supporting Information). Young's modulus
\textit{E} is the slope of the strain-stress curve. A large \textit{E}
means a rigid material. The calculated \textit{E} by linear fitting
equals 5.85~N/m, indicating a soft mechanical property of super-B. We also
verify this value by using elastic tensor
formula\cite{gao2017novel,gao2018unusually,wang2019ultralow}
\textit{E}=(C$_{11}$C$_{22}$-C$_{12}$C$_{21}$)/C$_{22}$. This ultrasoft
elastic property of super-B may have great potential for designing
flexible electronic devices\cite{zhang2017elasticity,wang2016strain}.

\begin{figure}[t!]
\includegraphics[width=1.0\columnwidth]{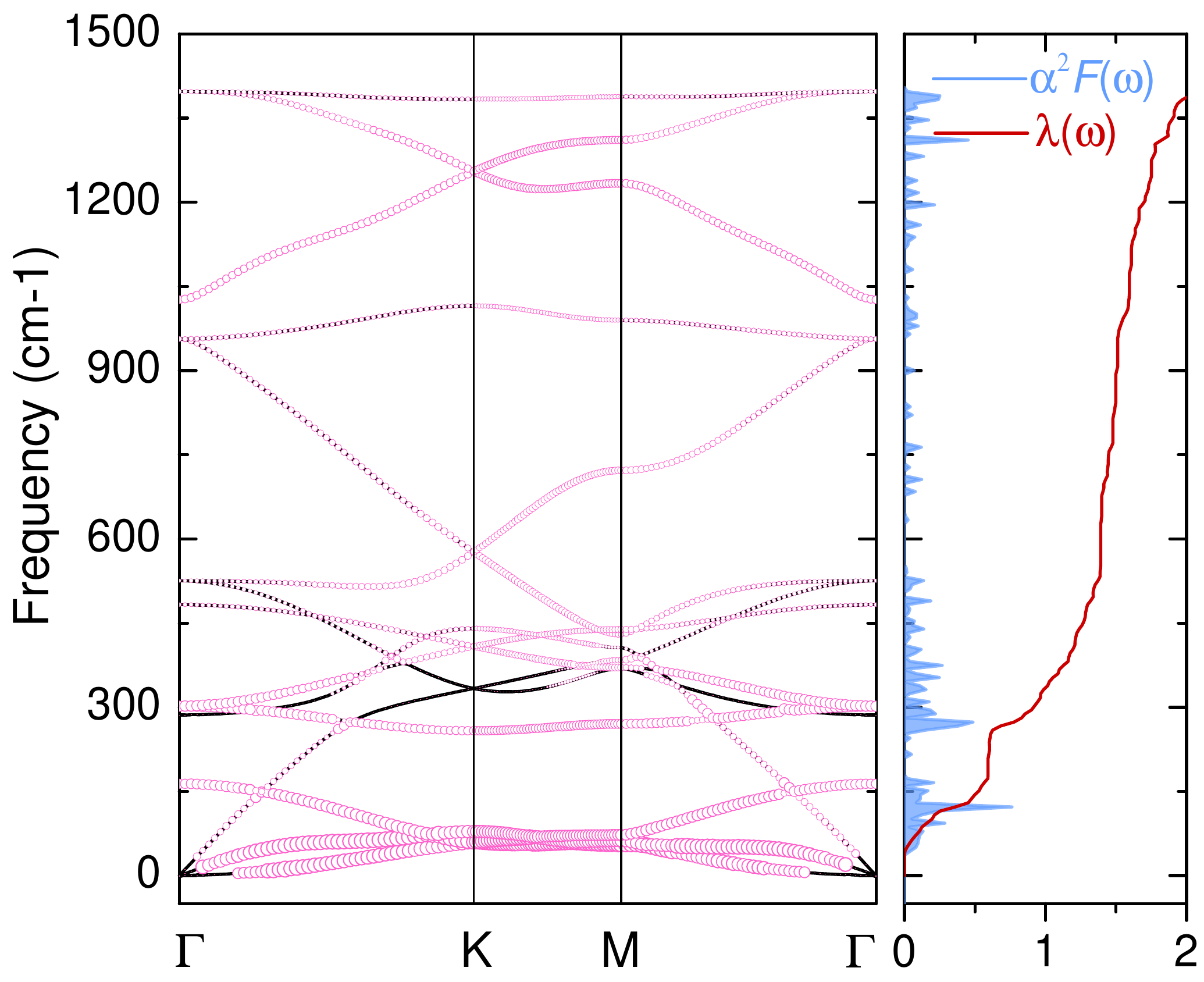}
\caption{Phonon band structure and electron-phonon coupling (EPC)
of the super-borophene. The area of the pink circle is proportional
to the EPC strength. The right panel is Eliashberg spectral function
\textit{$\alpha^2$F($\omega$)} and the total EPC constant
\textit{$\lambda(\omega)$}.}\label{fig6}
\end{figure}

\begin{table}
 \begin{threeparttable}
		\caption{Single-element superconductivity of graphene, silicene, phosphorene, stanene,
                 $\chi_3$-, $\beta_{12}$-, $\delta_6$- and super-borophene.}\label{Tab3}
		\renewcommand\arraystretch{1.5}
		\begin{tabular*}{0.50\textwidth}{p{2.3cm}p{1.1cm}p{2.1cm}p{2.0cm}}		
			\hline \hline
	        Materials           & ~~$\lambda$ & $\omega_\textrm{log}$  (cm$^{-1}$) & \textit{T$_c^{\mu^*=0.1}$} (K)    \\
			\hline
			Graphene$^a$        & 0.61   & ~~277.8   & ~~~8.1            \\
            Silicene$^b$        & 0.44   & ~~236.9   & ~~~1.7            \\
            Phosphorene$^c$     & 0.54   & ~~176.3   & ~~~4.2            \\
            Stanene$^d$         & 0.65   & ~~42.3    & ~~~1.3            \\
			$\chi_3$-B$^e$  	& 0.62   & ~~455.4   & ~~~11.5            \\
			$\beta_{12}$-B$^e$  & 0.78   & ~~362.0   & ~~~16.1      	    \\
			$\delta_6$-B$^e$  	& 1.05   & ~~272.5   & ~~~20.5            \\
			Super-B$^f$  	    & 1.95   & ~~102.1   & ~~~20.8            \\
			\hline \hline
		\end{tabular*}
 \begin{tablenotes}
        \footnotesize
        \item[a]Li doped graphene\cite{profeta2012phonon}. $^b$Silicene under 0.44 e/atom doping\cite{wan2013phonon}.
        $^c$Phosphorene under 0.10 e/atom doping\cite{shao2014electron}. $^d$Li doped stanene\cite{shaidu2016first}.
        $^{e}$Three experimental 2D borophenes\cite{penev2016can}. $^{f}$Present work.
      \end{tablenotes}
    \end{threeparttable}
\end{table}

The Poisson's ratio $\nu$, defined as the negative sign ratio of
lateral to applied strain\cite{gao2018two,gao2017novel}, is shown
in Figure~\ref{fig5}a in a pink curve. The equilibrium super-B
has a very large $\nu$ of 0.95, more than 5~times of borophene
with vacancy =1/8\cite{zhang2017elasticity} and also 5~times of
graphene\cite{liu2007ab}. This ultrahigh of $\nu$ suggests that
the response deformation of super-B is almost with the same
amplitude when applied strain to the pristine structure.
As strain increases, $\nu$($\varepsilon$) decreases significantly,
indicating a gradual weakness of Poisson effect at large strain.
The minimum $\nu$($\varepsilon$) is 0.03 at $\varepsilon$=24\%,
reducing to a final saturation near the critical
point\cite{liu2007ab,gao2018two}.


\subsection*{Superconductivity}
As an accepted rule of thumb, a good phonon-mediated superconductor
may satisfy some of conditions: (i) a small average atomic mass
\textit{\={M}}, (ii) a large electronic density of states (DOS) at
\textit{E$_F$}, (iii) a large Debye temperature $\theta_D$. As
$\theta_D$ is inversely proportional to \textit{\={M}}, a small
\textit{\={M}} generally means a large $\theta_D$. In this
sense, hydrogen is a good candidate for superconductor. However,
hydrogen is an insulator at normal condition, having a very small
DOS at \textit{E$_F$}. Metallic boron, also with very small
\textit{\={M}}, is a promising candidate to satisfy all above
criterion simultaneously. The famous bulk MgB$_2$ in which
magnesium is inserted into hexagonal boron, has an unprecedented
high \textit{T$_c$} of 39~K\cite{nagamatsu2001superconductivity}.

\begin{figure*}[t!]
\includegraphics[width=2.0\columnwidth]{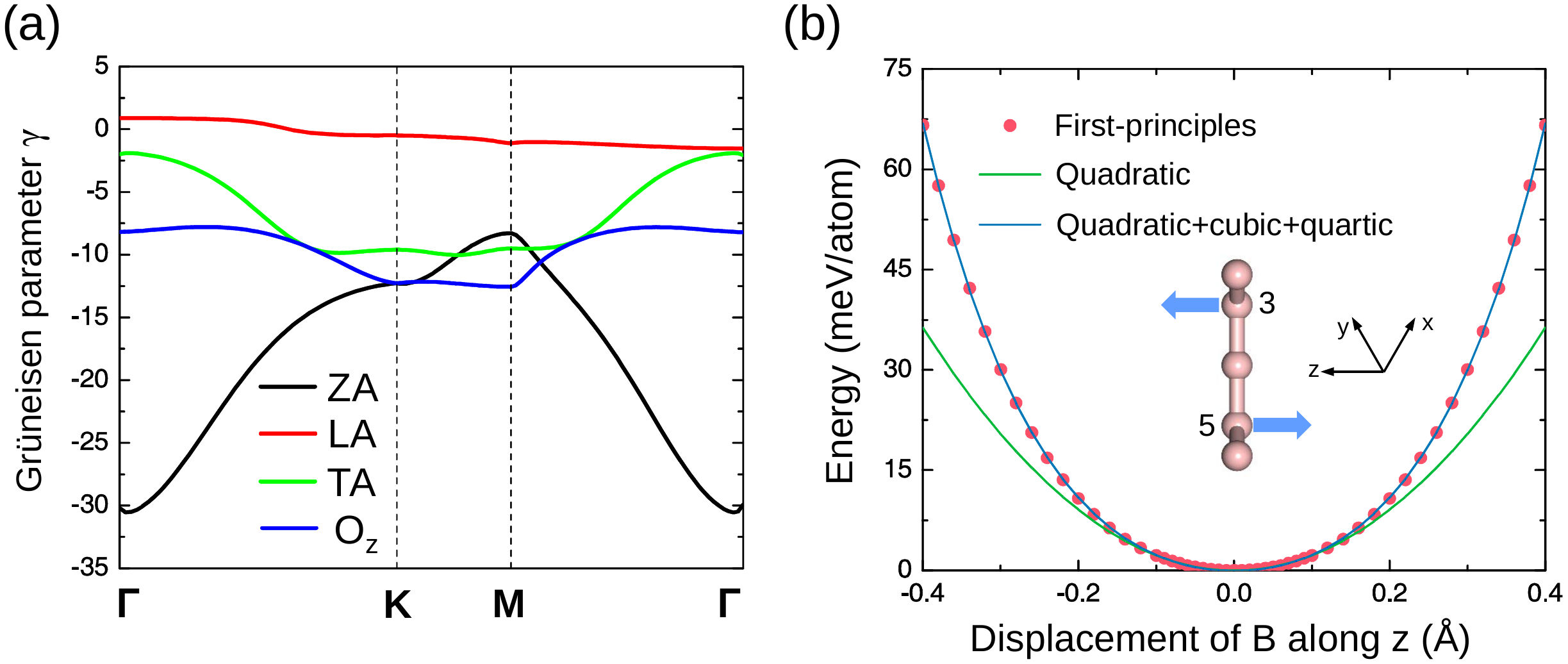}
\caption{(a) Gr\"{u}neisen parameter $\gamma$ of three acoustic and
one optical phonon branches in the first Brillouin zone. (b) Total
energy as a function of displacement of \textit{O$_z$} phonon mode,
in which B$_1$, B$_2$, and B$_4$, shown in Figure~\ref{fig1}a, are
static but B$_3$ and B$_5$ make movements in an opposite direction
as an inset.
}\label{fig7}
\end{figure*}

According to the Bardeen–Cooper–Schrieffer (BCS) theory and Eliashberg
equations, the spectral function can be expressed as\cite{bardeen1957microscopic}
\begin{equation} %
\label{eq1}
\alpha^2F(\omega) = \frac{1} {2 \pi N(E_F)} \sum_{\textbf{q}\nu} \delta(\omega-\omega_{\textbf{q}\nu}) \frac{\gamma_{\textbf{q}\nu}} {\hbar \omega_{\textbf{q}\nu}},
\end{equation}
where \textit{N}, $\omega_{\textbf{q}\nu}$, and $\gamma_{\textbf{q}\nu}$
are the electronic density of states at \textit{E$_F$}, phonon frequency
and linewidth for phonon modes $\lambda$ with the wave vector \textbf{q}.
The total EPC constant can be obtained when making a summation of spectral
function over the first Brillouin zone
\begin{equation} %
\label{eq2}
\lambda = \sum_{\textbf{q}\nu} \lambda_{\textbf{q}\nu} = 2 \int \frac{\alpha^2F(\omega)} {\omega} d\omega,
\end{equation}
Finally, based on the Allen–Dynes formula\cite{allen1975transition},
the critical temperature \textit{T$_c$} of superconductor can be
estimated by
\begin{equation} %
\label{eq3}
T_c = \frac{\omega_\textrm{log}} {1.2} \textrm{exp} [\frac{-1.04(1+\lambda)} {\lambda(1-0.62\mu^*)-\mu^*}],
\end{equation}
in which $\mu^*$ is the effective screened Coulomb repulsion constant
(generally 0.10$\sim$0.15) and $\omega_\textrm{log}$ reads
\begin{equation} %
\label{eq4}
\omega_\textrm{log} = \textrm{exp} [\frac{2} {\lambda} \int \frac{d \omega} {\omega} \alpha^2F(\omega) \textrm log(\omega)].
\end{equation}

The calculated phonon dispersion, EPC, and  spectral function of super-B
are shown in Figure~\ref{fig6}. The area of pink circle is proportional
to the strength of EPC.
The largest EPC strength stems from the ZA, TA and one special optical
phonon mode.
For the purely flat 2D materials, such as graphene, ZA
mode must satisfy a symmetry-based selection
rule\cite{lindsay2010flexural}. Therefore, for 2D materials, ZA
modes generally dominant the heat transport and show a giant
anharmonicity (large Gr\"{u}neisen parameter).

As the vibrational direction of this optical phonon is
parallel to the \textit{z} axis, we call it \textit{O$_z$} mode for
the sake of simplicity.
The \textit{O$_z$} mode, anomalously, drops into the acoustic phonon
regions which is in spirit similar with recent discovered 2D
tellurene\cite{gao2018unusually}. The low optical
\textit{O$_z$} mode makes it easier to satisfy the energy and
momentum conservation simultaneously, leading to a large phase
space of phonon-phonon scattering\cite{gao2018unusually}.
The EPC strength of this \textit{O$_z$}
mode firstly increases from $\Gamma$ to K, and M and then decreases when
approaching $\Gamma$ again.

The corresponding Eliashberg spectral function \textit{$\alpha^2$F($\omega$)},
in Figure~\ref{fig6}, also reveals that this \textit{O$_z$} mode enhances the
EPC in the low-frequency of acoustic phonon region. The total EPC constant
$\lambda$ and critical temperature \textit{T$_c$}, according to the
Eq. \eqref{eq1}-\eqref{eq4}, are shown in Table ~\ref{Tab3}. The
phonon-mediated \textit{T$_c$} of super-B is 20.8~K with
$\lambda$=1.95 (at $\mu$=0.1). We also list some other related 2D
materials and borophene allotropes. Our super-B has the
first-class intrinsically critical temperature \textit{T$_c$} of 20.8~K,
which is comparable and even a little higher than the $\delta_6$-B (20.5~K).
To the best of knowledge, super-B has the highest intrinsically critical
temperature \textit{T$_c$} in single-element superconductors at ambient condition.

Furthermore, Profeta et al.\cite{profeta2012phonon} proposed a seminal strategy
to significantly enhanced the \textit{T$_c$} of LiC$_6$ by inserting the metal
lithium due to a large density of state at Fermi level. We hope \textit{T$_c$}
of super-B can also be further enhanced by a similar method, such as metal
potassium\cite{kambe2019solution}.

After analysis of separated phonon modes, we find that the high
\textit{T$_c$} of super-B mainly stems from the giant phonon
anharmonicity.
ZA, TA and \textit{O$_z$} phonon modes are the three main
contributions to the large EPC, and finally large \textit{T$_c$}.
The Gr\"{u}neisen parameter $\gamma$, in Figure~\ref{fig7}a,
shows volume response as a function of phonon thermal vibration
in super-B. A large $\gamma$ means a strong anharmonicity in chemical
bonds of materials\cite{gao2018unusually,gao2016heat,gao2016stretch}.
The largest value of $\gamma$ is around $-30$, belonging to the ZA mode
in black line. The secondary is \textit{$O_z$} branch in blue line
having an average value of $-10$. This unusually low optical phonon
branch would enhance the phonon-phonon scattering, as well as
electron-phonon scattering.
Notably, this is also evidenced by the fact that the
soft phonon modes are eliminated with applied strain in Figure~\ref{fig5}b.
Specifically, the strong acoustic-acoustic phonon and acoustic-optical
phonon interactions lead to a large phonon anharmonicity. This giant
anharmonicity, finally, brings about the large peak in the Eliashberg spectral
function \textit{$\alpha^2$F($\omega$)} in the right panel in Figure~\ref{fig6}.
Besides, based on the Eq. \eqref{eq2}, we obtain that there is 60\% $\lambda$
below 300 cm$^{-1}$ in the total EPC contributions, which verifies that
low frequency significantly is responsible for this unusually large $\lambda$.

To further explore the phonon anharmonic property in super-B, we
calculate the total energy as a function of displacement in
phonon \textit{O$_z$} mode, in which B$_1$, B$_2$, and B$_4$
shown in Figure~\ref{fig1}a are static but B$_3$ and B$_5$ make
movements in an opposite direction, in Figure~\ref{fig7}b. ZA
and TA modes also have similar behaviors. The
first principle results can be well fitted by a polynomial
expression up to quartic
\textit{E(u)}=225.56\textit{u$^2$}+1203.01\textit{u$^4$}
in the blue line. However, harmonic approximation (green line)
cannot produce a good agreement. The ratio between quartic and
quadratic terms is \textit{A$_4$}/\textit{A$_2$}$\approx$5.33, indicating
a quite large anharmonicity in super-B, which finally leads to a
strong interaction between electrons and phonons. To the best of
knowledge, superconductor MgB$_2$ binary alloy\cite{yildirim2001giant}
and 2D tellurene with high figure of merit
\textit{zT}\cite{gao2018unusually,gao2018high} also exhibit very
large phonon anharmonicity.

For the band structure, we consider the distorted and undistorted
super-B due to the \textit{O$_z$} phonon mode. The result is shown in the
Supporting Information. The band splitting between the undistorted and
distorted structures in super-B is around 0.07~eV. Although this number
is smaller than the MgB$_2$ (\textit{T$_c$}=39~K)\cite{nagamatsu2001superconductivity},
its impact is significantly large. We mark a special area with a blue
circle which is close to the Fermi level and more importantly, and its DOS
is quite large due to the relatively flat band. Hence, giant anharmonicity
of this \textit{O$_z$} phonon mode intensifies the DOS around the Fermi
level, then stimulates the EPC and finally leads to a high \textit{T$_c$}.

\section*{Discussion}
Similar to carbon nanotube, boron can also be rolled into a
cylinder\cite{tang2007novel,wu2012two}. We optimize
singled-walled borophene nanotubes (SWBNTs) with different
diameters and edge shapes (armchair or zigzag). We find that
all SWBNTs are quite stable remaining a good cylinder without
any buckling or collapse. Results are shown in the table of
the Supporting Information.
The most important result is that SWBNTs are always good metal. In
order to further confirm the metallic property, for example,
we plot the DOS of zigzag (8,0) and armchair (8,8) tubes in the
Supporting Information. At the \textit{E$_F$}, DOS is significantly
enhanced when rolling into a cylinder, which is irrelevant to
the type of edge shapes. This probably will, in turn, further
amplify the \textit{T$_c$} of SWBNTs according to the rule (ii)
mentioned above. According to the previously seminal
works\cite{benedict1995static}, 
\textit{T$_c$} will be unquestionably enhanced by rolling sheets
into tubes as a result of opening new electron-phonon scattering
channels. However, due to formidable computing requirement, currently,
one is quite difficult to simulate the EPC in SWCNTs and SWBNTs based
on the accurate first-principle calculations. As one of the nearest
neighbors of carbon, super-B and SWBNTs are promising superconductors by
using the similar method of
carbon\cite{profeta2012phonon}.

Compared with $\chi_3$-, $\beta_{12}$-, and $\delta_6-$B, super-B is a
metastable phase of borophene. In retrospect, the most stable phase of
2D silicon is silicene with 18 atoms in the primitive
cell\cite{vogt2012silicene}, rather than the phase with 2
atoms\cite{cahangirov2009two}. Even though the latter is much less stable
than the previous, the latter, at present, could be easily grown with very
accurate experimental technique\cite{fleurence2012experimental,feng2012evidence}.
Similarly, blue phosphorus\cite{zhu2014semiconducting}, as a metastable of black
phosphorus\cite{liu2014phosphorene}, has also been successfully grown on
Au(111)\cite{gu2017growth,zhang2016epitaxial}. The cohesive energy \textit{E$_c$} of
the experimentally attainable silicene with 2 atoms and phosphorene
are 3.71~eV/atom\cite{fleurence2012experimental,feng2012evidence}
and 3.61~eV/atom\cite{liu2014phosphorene,li2014black}, separately.
A large \textit{E$_c$} indicates a strong chemical bond
in materials. The \textit{E$_c$} of super-B is 5.55~eV, indicating
a strong and robust chemical bond to maintain stability. Furthermore,
due to the complex chemical environment and variable substrate effect,
a recently discovered borophene has been synthesised\cite{mannix2015synthesis}.
Notably, it is a novel metastable borophene, also less stable than
the $\chi_3$-, $\beta_{12}$-, and $\delta_6-$B allotropes.

Very recently, a cutting-edge experimental technique, called reactive
molecular beam epitaxy method, has been successfully used to attain the
freestanding crystalline oxide perovskites\cite{ji2019freestanding}.
Borophene, like oxide perovskites, has no counterpart in bulk material.
Hence, except for the metal substrates, this recently developed approach
may shed light on the experimental realization of super-B and other
boron-related nanostructures.

Furthermore, a 2D atomic layer with hexagonal boron network
has been bottom-up synthesized freshly. The main structure in their
work\cite{kambe2019solution} is exactly our predicted super-B, which
further verifies the correctness of our theoretical calculation and
the importance of our work.

As a matter of fact, for strong coupling systems ($ \lambda >1.5 $),
superconducting temperature \textit{T}$_c$ should be calculated using a
more general expression according to Allen and Dynes\cite{allen1975transition}
\begin{equation} %
\label{Tcc}
T_c = \frac{f_1 f_2 \omega_{log}} {1.2} \textrm{exp} [- \frac{1.04(1+\lambda)} {\lambda-\mu^* (1+0.62\lambda)} ],
\end{equation}
\begin{equation} %
\label{f1}
f_1 = [1 + (\lambda/\Lambda_1)^{3/2} ]^{1/3},
\end{equation}
\begin{equation} %
\label{f2}
f_2 = 1 +  \frac{ (\bar \omega_2/\omega_{log} -1) \lambda^2 } {\lambda^2 + \Lambda_2^2},
\end{equation}
where \textit{f}$_1$ and \textit{f}$_2$ represent the strong-coupling correction and
shape correction, respectively. The parameters $\Lambda_1$ and $\Lambda_2$ are given
by\cite{allen1975transition}
\begin{equation} %
\label{para1}
\Lambda_1 = 2.46(1+3.8\mu^*),
\end{equation}
\begin{equation}
\label{para2}
\Lambda_2 = 1.82(1+6.3\mu^*)(\bar \omega_2/\omega_{log}),
\end{equation}
in which the general moment is defined as\cite{allen1975transition}
\begin{equation}
\label{moment}
 <\omega^n>= \frac{2} {\lambda} \int d\omega \alpha^2 F(\omega) \omega^{n-1},
\end{equation}

According to Eq. \eqref{Tcc}-\eqref{moment}, the calculated \textit{T}$_c$ is
25.3~K when $\mu^*=0.1$. Note that this result is larger than 20.8~K based
on Eq. \eqref{eq3}, further confirming our conclusion that super-B has the
highest critical temperature \textit{T}$_c$ in single-element superconductors
at ambient conditions.

In summary, we have predicted a previously unknown monolayer borophene
by first principles.  It has good thermal, dynamical, and mechanical
stability compared with many other typical borophenes. We have found
that super-B has a fascinating chemical bond environment consisting
of standard \textit{sp}, \textit{sp$^2$} hybridizations and delocalized
five-center three-electrons $\pi$(5c-3e) bonds,
based on the
NBO analysis. This exceptional electronic structure plays a
crucial role in stabilizing the super-B chemically. Meanwhile, By extra
doping, super-B can be transformed into a Dirac material from pristine
metal. Like graphene, it also has a superior flexibility. Furthermore,
due to the small atomic mass and large density of state at Fermi level,
super-B has the highest intrinsically critical temperature \textit{T$_c$}
of 25.3~K in single-element superconductors at ambient condition. We have
attributed this high \textit{T$_c$} of super-B to the giant anharmonicity
of two linear acoustic phonon branches and an unusually low optic
\textit{O$_z$} phonon mode.



\quad\\
{\noindent\bf Author Information}\\

{\noindent\bf Corresponding Author}\\
$^*$E-mail: {\tt zhibin.gao@nus.edu.sg} \\

{\noindent\bf ORCID}\\
Zhibin Gao: 0000-0002-6843-381X \\


{\noindent\bf Supporting Information}\\
The Supporting Information is available free of charge on the
ACS Publications website via the Internet at
https://pubs.acs.org/journal/aamick.

Phonon dispersion of super-B, electronic band structures of
pristine $\alpha$-graphyne and super-B under $-0.2$ e/B doping,
and movies of molecular dynamics simulations at T=600~K at the
end of 10~ps.\\

{\noindent\bf Notes}\\
The authors declare no competing financial interest.


\begin{acknowledgement}\\
We acknowledge David Tom\'{a}nek for many fruitful discussions and
good suggestions.We also thank Hanyu Liu for valuable discussions
and kind help. M.L. acknowledges useful discussions with Xiang Zhao.
The HPC platform of Xi'an Jiaotong University is highly appreciated.
This work is supported by an MOE tier 1 grant R-144-000-402-114.
\end{acknowledgement}

%

\providecommand{\latin}[1]{#1}
\makeatletter
\providecommand{\doi}
  {\begingroup\let\do\@makeother\dospecials
  \catcode`\{=1 \catcode`\}=2 \doi@aux}
\providecommand{\doi@aux}[1]{\endgroup\texttt{#1}}
\makeatother
\providecommand*\mcitethebibliography{\thebibliography}
\csname @ifundefined\endcsname{endmcitethebibliography}
  {\let\endmcitethebibliography\endthebibliography}{}

\end{document}